\documentclass[conference,a4paper]{IEEEtran}

\usepackage[utf8]{inputenc} 
\usepackage[T1]{fontenc}
\usepackage{url}              
\usepackage{cite}             
\usepackage{amssymb}
\usepackage[cmex10]{amsmath}  
\usepackage{mleftright}       
\mleftright                   

\usepackage{graphicx}         
\usepackage{booktabs}         

\usepackage{algorithm} 
\usepackage{algorithmic}

\newtheorem{theorem}{Theorem}[section]
\newtheorem{proposition}[theorem]{Proposition}

\newcommand{\blds}[1]{\boldsymbol{#1}}
\newcommand{\opr}[1]{\operatorname{#1}}

\begin{document}

\title{The Asymptotics of Difference Systems of Sets for Synchronization and Phase Detection} 

\author{%
  \IEEEauthorblockN{Yu Tsunoda}
  \IEEEauthorblockA{Faculty of Engineering, Information and Systems\\
                    University of Tsukuba\\
                    Email:  y.tsunoda@sk.tsukuba.ac.jp}
  \and
  \IEEEauthorblockN{Yuichiro Fujiwara}
  \IEEEauthorblockA{Division of Mathematics and Informatics\\
                    Chiba University\\ 
                    Email: yuichiro.fujiwara@chiba-u.jp}
}

\maketitle

\begin{abstract}
We settle the problem of determining the asymptotic behavior of the parameters of optimal difference systems of sets, or DSSes for short, which were originally introduced for computationally efficient frame synchronization under the presence of additive noise.
We prove that the lowest achievable redundancy of a DSS asymptotically attains Levenshtein's lower bound for any alphabet size and relative index,
answering the question of Levenshtein posed in 1971.
Our proof is probabilistic and gives a linear-time randomized algorithm for constructing asymptotically optimal DSSes with high probability for any alphabet size and information rate.
This provides efficient self-synchronizing codes with strong noise resilience.
We also point out an application of DSSes to phase detection.
\end{abstract}

\section{Introduction}
\label{sec:intro}

Synchronization between the sender and receiver is a fundamental problem in communications.
Because data cannot be parsed without proper synchronization, it is among the oldest yet still critical problems in information theory \cite{Golomb:1963aa}.

The kind of synchronization we consider in this paper is frame synchronization, which addresses detection of the boundaries of each information block under the assumption that synchronization at the symbol level is already provided.
For instance, the reader familiar with the Latin alphabet can recognize each letter in this sentence, which means that the knowledge of the alphabet provides synchronization at the symbol level.
However, frame synchronization in English text is mainly achieved by spaces between words; without them, we cannot tell, in principle, whether ``thisisinformation'' should be understood as ``this is information'' or ``this is in formation.''

As is the case with many other fundamental problems in information theory, solutions to frame synchronization have long been investigated in various contexts, including well-known prefix codes \cite{Kraft:1949aa} and Huffman codes \cite{Huffman:1952aa} for data compression, comma-free codes \cite{Golomb:1958aa,Crick:1957aa,Eastman:1965aa} in the context of noiseless communications, and, as an example of more recent topics, quantum synchronizable codes \cite{Fujiwara:2013,Fujiwara:2013d,Fujiwara:2014aa,Xie:2016aa,Luo:2018aa,Li:2019aa,Du:2020aa,Li:2021aa,Dinh:2021aa} for noisy quantum information.
Among many important questions in this field, in this paper our particular interest lies in a coding-theoretic problem of frame synchronization for digital data transmission under the presence of additive noise \cite{Golomb:1963aa,Scholtz:1980aa}.

To formally describe codes of our interest, let $q \geq 2$ and $n \geq \rho \geq 0$ be integers.
Take a finite set $A_q$ of cardinality $q$.
For an ordered pair of sequences $\boldsymbol{x} = (x_0,\dots,x_{n-1}), \boldsymbol{y} = (y_0,\dots, y_{n-1}) \in A_q^n$ of length $n$ over $A_q$, a \textit{splice} of $\boldsymbol{x}$ and $\boldsymbol{y}$ is a concatenated sequence $ (x_{t}, \dots, x_{n-1},y_0,\dots, y_{t-1})$ of the last $n-t$ digits of $\boldsymbol{x}$ and the first $t$ digits of $\boldsymbol{y}$ for some integer $t$, $1 \leq t \leq n-1$. A $q$-\textit{ary self-synchronizing code} of \textit{length} $n$ and \textit{comma-free index} $\rho$ over $A_q$ is a set $\mathcal{C} \subseteq A_q^n$ of sequences, called \textit{codewords}, of length $n$ such that the Hamming distance between any codeword $\boldsymbol{z} \in \mathcal{C}$ and any splice of any pair of codewords is at least $\rho$. A concise introduction to this type of code and relevant recent results can be found in \cite{Levenshtein:2004,Fujiwara:2013h}.

A self-synchronizing code of comma-free index $1$ is more commonly known as a \textit{comma-free code}, where the requirement is simply that no splices of two codewords form a valid codeword.
Assuming data is transmitted in the form of consecutive codewords through a noiseless channel, this property allows for detecting the boundaries between adjacent codewords in the stream of digits \cite{Golomb:1958aa}. Indeed, a contiguous subsequence forms a valid codeword only when it is properly aligned so that it starts and ends at correct boundaries.

In general, a self-synchronizing code of length $n$ and comma-free index $\rho$ allows for recovering from frame misalignment under the assumption that there are at most $\lfloor\frac{\rho-1}{2}\rfloor$ symbol substitution errors within any $n$ consecutive digits because the Hamming balls of radius $\lfloor\frac{\rho-1}{2}\rfloor$ centered around codewords share no elements with those of the same radius around splices of codewords \cite{Levenshtein:1969aa}. In other words, comma-free index is to frame synchronization as minimum distance is to error correction in algebraic coding theory \cite{MacWilliams:1977}. However, this entails the fact that having nontrivial comma-free index alone is not enough to realize computationally efficient encoding and decoding of a self-synchronizing code.

A difference system of sets, or DSS for short, is a combinatorial solution to this problem, which was first introduced in \cite{Levenshtein:1971}.
As will be defined formally in the next section, it is a special combinatorial design that cleverly inserts synchronization markers into the codewords of any efficient block error-correcting code $\mathcal{C}$ to form a self-synchronizing code that directly exploits fast encoding and decoding algorithms for $\mathcal{C}$.

A DSS has four basic parameters, which we shall denote here by letters $q$, $n$, $r$, and $\rho$ and refer to a specific DSS as a $q$-ary DSS$(n,r,\rho)$ to spell out these parameters. When combined with a $q$-ary code $\mathcal{C}$, the resulting self-synchronizing code is $q$-ary, of length $n$, and of comma-free index $\rho$. The parameter denoted by $r$ represents the number of digits used as synchronization markers, so that the $n-r$ remaining digits in a codeword may be used to carry information. For this reason, this parameter $r$ is called the \textit{redundancy} of a DSS.

All else being equal, we would like the redundancy of our DSS to be as small as possible. The following is the most fundamental lower bound on the redundancy of a DSS. 
\begin{theorem}[Levenshtein bound \cite{Levenshtein:1971}]\label{thm:levenshtein}
Let $q \geq 2$ and $n \geq \rho \geq 1$ be integers. For any $q$-ary DSS$(n,r,\rho)$, it holds that
\begin{align*}
r \geq \sqrt{\frac{q\rho(n-1)}{q-1}}.
\end{align*}
\end{theorem}

For given $q$, $n$, and $\rho$, let $r_q(n,\rho) = \min\{r \in \mathbb{N}_0 \mid \text{there exists a $q$-ary DSS$(n,r,\rho)$}\}$ be the smallest possible redundancy of a $q$-ary DSS for given $n$ and $\rho$.
From a coding-theoretic point of view, we are interested in the trade-off between \textit{redundancy rate} $\frac{r}{n}$ or, equivalently, \textit{information rate} $1-\frac{r}{n}$, and \textit{relative index} $\delta = \frac{\rho}{n}$. In this context, a fundamental question is whether $\frac{r_q(n,\lceil n\delta\rceil)}{n}$ is bounded away from $1$ for positive constant $\delta$ as $n \rightarrow \infty$. In view of Theorem \ref{thm:levenshtein}, we may even ask whether $\frac{r_q(n,\lceil n\delta\rceil)}{n} \sim \sqrt{\frac{q\delta}{q-1}}$ for fixed $q$ and $\rho$ as $n \rightarrow \infty$. If true, the mature theory of error correction can be imported wholesale with only a little sacrifice in rate \cite{Levenshtein:2004}.

Unfortunately, our knowledge in the literature regarding the asymptotic behavior of rate and relative index is quite limited.
Still, Theorem \ref{thm:levenshtein} is known to be asymptotically tight for binary DSSes with vanishing relative index.
\begin{theorem}[\!\!\cite{Levenshtein:1971}]\label{thm:vanish}
Let $\rho$ be a positive integer function of $n$ such that $\frac{\rho}{n} \rightarrow 0$ as $n \rightarrow \infty$. Then, it holds that
$
r_2(n,\rho) \sim \sqrt{2\rho n}
$
as $n$ tends to infinity.
\end{theorem}

Several infinite sequences of DSSes with vanishing information rate are also known to attain the lower bound in Theorem \ref{thm:levenshtein} \cite{Chang:2006,Ding:2008,Fan:2008aa,Chee:2010,Zhou:2010aa,Zhou:2012,Cai:2013aa,Ding:2014aa,Zha:2015aa,Cai:2017aa,Yi:2019aa}.
However, for the most important case of non-vanishing information rate and non-vanishing relative index, the problem of determining asymptotic behavior of the parameters of best possible DSSes has remained completely open despite the numerous, albeit somewhat sporadic, discoveries of interesting DSSes such as those found in \cite{Tonchev:2005,Fuji-Hara:2006,Tonchev:2007,Mutoh:2008,Fuji-Hara:2009,Lei:2011aa,Fan:2012,Fujiwara:2013h,Fujiwara:2013a,Fan:2014aa,Chisaki:2015aa,Qi:2017aa,Chisaki:2019aa,Wang:2020aa}.

The primary purpose of this paper is to settle this asymptotic problem in its entirety.

\begin{theorem}\label{thm:main}
For any integer $q \geq 2$ and any constant $p \in (0,1)$, there exists a constant $n_p$ such that for any integer $n \geq n_p$, there exists a $q$-ary DSS$(n,\lfloor np\rfloor,\rho)$, where
\begin{align*}
\rho = n\left(1-\frac{1}{q}\right)p^2 - o(n).
\end{align*}
In particular, for any constant integer $q \geq 2$ and any constant $\delta \in \left(0,1-\frac{1}{q}\right)$,
\begin{align*}
\lim_{n \rightarrow \infty} \frac{r_q(n,\lceil n\delta\rceil)}{n} = \sqrt{\frac{q\delta}{q-1}}.
\end{align*}
\end{theorem}

Because the case where relative index $\delta \geq 1-\frac{1}{q}$ is vacuous, the above theorem completely solves the asymptotic problem.

Another important purpose of this paper is to show that $q$-ary DSSes with strong noise resilience can be efficiently constructed for any desired information rate.
In fact, we prove Theorem \ref{thm:main} by giving a randomized algorithm that runs in time linear in $n$ and produces with high probability DSSes that asymptotically attain Theorem \ref{thm:levenshtein}. While our results are asymptotic in nature, this algorithmic aspect may be of equal importance when DSSes need to be explicitly constructed for use in applications proposed in the literature, such as pulse position modulation \cite{Fujiwara:2013a} and authentication and secrete sharing schemes \cite{Ogata:2004}.

Finally, we give a new application of efficient self-synchronizing codes to the problem of phase detection \cite{Wang:2017aa}. Here, instead of aiming for precisely attaining bounds on the parameters of codes for phase detection, we  show that phase detection schemes can be realized with a variety of computationally efficient encoders and decoders known in the theory of error correction, illustrating the versatility of DSSes.

In the next section, we briefly review difference systems of sets and a few algorithmic facts relevant to our results. Section \ref{sec:proof} gives the proof of our main result, namely Theorem \ref{thm:main}. We conclude this paper in Section \ref{sec:conclusion} with a few remarks, where an application to phase detection is also touched on.

\section{Preliminaries}\label{sec:pre}

In this section, we briefly review relevant facts and notions in combinatorics and algorithms. Section \ref{subsec:dss} formally defines a difference system of sets and relates this concept to self-synchronizing codes. Section \ref{subsec:random} deals with algorithmic random permutations \cite{Skiena:2020aa} 
and their use in the context of concentration of measure in large deviation theory \cite{McDiarmid:1989aa,Talagrand:1995aa}.

\subsection{Difference Systems of Sets}\label{subsec:dss}

Formally, a $q$-\textit{ary} \textit{difference system of sets} (DSS) of \textit{index} $\rho$ over the cyclic group $\mathbb{Z}_n$ of order $n$ is a family of $q$ disjoint subsets $Q_0, \dots, Q_{q-1} \subseteq \mathbb{Z}_n$ such that the multiset $\{a-b \mid a \in Q_i, b \in Q_j, i\not=j\}$ of \textit{external differences} contains each nonzero element of $\mathbb{Z}_n$ at least $\rho$ times.
The cardinality $r = \lvert \bigcup_{i=0}^{q-1} Q_i \rvert$ is called the \textit{redundancy} of the DSS.
A DSS of index $\rho$ and redundancy $r$ over $\mathbb{Z}_n$ is denoted by DSS$(n,r,\rho)$.
In what follows, we always identify the elements of $\mathbb{Z}_n$ with the integers modulo $n$ so that $\mathbb{Z}_n = \{0,1,\dots,n-1\}$.

To see how a $q$-ary DSS$(n,r,\rho)$ gives a self-synchronizing code, take a set $A_q = \{a_0,\dots,a_{q-1}\}$ of cardinality $q$ and another distinct element $*  \not\in A_q$ to form a set $A_q \cup \{*\}$ of $q+1$ elements.
Define $h \colon (A_q \cup \{*\}) \times (A_q \cup \{*\}) \rightarrow \{0,1\}\subseteq\mathbb{N}_0$ by
\begin{align*}
h(x,y) = 
\begin{cases}
1 & \text{if $x \not= y$ and $x,y \in A_q$,}\\
0 & \text{otherwise.}
\end{cases}
\end{align*}
We construct a $q$-ary self-synchronizing code of length $n$ and comma-free index $\rho$ over $A_q$ by combining an arbitrary $q$-ary error-correcting code $\mathcal{C}$ of length $n-r$ with a sequence $\boldsymbol{v} = (v_0,\dots,v_{n-1})$ of length $n$ over $A_q\cup\{*\}$ corresponding to a $q$-ary DSS$(n,r,\rho)$ $\{Q_0,\dots,Q_{q-1}\}$. The \textit{template sequence} $\blds{v}$ is defined by regarding each subset $Q_i$ as the set $\{j \in \{0,1,\dots,n-1\} \mid v_j = a_i\}$ of positions of $a_i \in A_q$ in $\boldsymbol{v}$ and assigning the extra element $*$ to the remaining $n-r$ positions specified by $\mathbb{Z}_n\setminus\bigcup_{i=0}^{q-1}Q_i$.
By replacing the $n-r$ $*$'s with each codeword of $\mathcal{C}$, we obtain a set $\mathcal{D}$ of $\vert \mathcal{C}\vert$ $q$-ary sequences.
Note that for any integer $t$, $1 \leq t \leq n-1$, we have
\begin{align*}
\sum_{i=0}^{n-1}h(v_i,v_{i+t}) \geq \rho
\end{align*}
because each external difference appears at least $\rho$ times in our DSS. Thus, the resulting set $\mathcal{D}$ is a self-synchronizing code of comma-free index $\rho$ regardless of the choice of $\mathcal{C}$. Because this approach completely separates frame synchronization from error correction on the payload, the resulting self-synchronizing code can exploit the encoding and decoding algorithms for $\mathcal{C}$ in the straightforward manner.

As a concrete example, let $Q_0 = \{1,2,3,4,6,15\}$ and $Q_1 =\{5,9,10,14,17,24\}$ over $\mathbb{Z}_{25}$. This pair forms a binary DSS$(25,12,3)$.
Our template sequence $\boldsymbol{v} = (v_0,\dots,v_{24})$ of length $25$ is defined by $v_i = j$ for $i \in Q_j$ and $v_i = *$ for $i \not\in Q_0\cup Q_1$, so that we have
\begin{align*}
{*}000010{*}{*}11{*}{*}{*}10{*}1{*}{*}{*}{*}{*}{*}1.
\end{align*}
These six $0$'s and six $1$'s are our synchronization markers. Placing the codewords of an arbitrary binary error-correcting code $\mathcal{C}$ of length $13$ on the free positions marked by $*$, we obtain a self-synchronizing code of comma-free index at least $3$ with $\vert \mathcal{C}\vert$ codewords. The payload is protected by $\mathcal{C}$ while we can detect frame misalignment by watching for excessive discrepancies in the twelve positions at which we should have synchronization markers as specified by the DSS.

\subsection{Random Permutations and Concentration of Measure}\label{subsec:random}

In what follows, the set $\{0,1,\dots,x-1\}$ for positive integer $x$ is denoted by $I_x$.
Let $\boldsymbol{v} = (v_0,\dots,v_{n-1})\in A_q^n$ be a sequence of length $n$ over a set $A_q$ of cardinality $q$.
The \textit{permutation} $\pi_{\sigma} \colon A_q^n \rightarrow A_q^n$ for $\boldsymbol{v}$ \textit{induced} by a permutation $\sigma \colon I_n \rightarrow I_n$ of its index set $I_n$ is defined by
$\pi_{\sigma}(\blds{v}) = (v_{\sigma(0)},\dots,v_{\sigma(n-1)})$. When $\sigma$ is a transposition of $I_n$, it is called the \textit{transposition induced} by $\sigma$.
An induced permutation is simply a reordering of a sequence, while an induced transposition just swaps a pair in positions in a sequence.

Let $\tau$ be a random variable whose codomain is the set of permutations of $I_n$, that is, the symmetric group $S_n$.
Then, the induced permutation $\pi_{\tau}$ may be seen as a random variable that shuffles a sequence $\blds{v}$ of length $n$, which we call the \textit{random permutation induced} by $\tau$.
When $\tau$ follows the uniform distribution on $S_n$, the corresponding induced random permutation $\pi_{\tau}$ is said to be \textit{uniform}.

Random permutations are fundamental in algorithms and can be seen as the opposite of sorting. Of particular interest to us is the well-known Fisher-Yates shuffle \cite{Skiena:2020aa}, of which the modern algorithmic equivalent by Durstenfeld \cite{Durstenfeld:1964aa} for computer implementation is known as the Knuth shuffle \cite{Knuth:1998aa}.

For positive integer $n$, let $t_0$, $t_1$, \dots, $t_{n-2}$ be $n-1$ mutually independent random variables such that for any $i \in I_{n-1}$ and any $j\in I_{n-i}$, it holds that $\opr{Pr}(t_i = j) = \frac{1}{n-i}$.
Consider $n-1$ random transpositions $\tau_0 = (n-1, t_0)$, $\tau_1 = (n-2, t_1)$, \dots, $\tau_{n-2} = (1, t_{n-2})$ of $I_n$, where $(a,b)$ in cycle notation represents swapping the elements $a$ and $b$.
Formally, the \textit{Knuth shuffle} is a randomized algorithm that takes a sequence $\blds{v}$ of length $n$ as its input and returns a reordered sequence $\pi_{\tau_{n-2}}\cdots\pi_{\tau_0}(\blds{v})$ of $\blds{v}$ by successively applying $\pi_{\tau_0}$, \dots, $\pi_{\tau_{n-2}}$, so that  $\pi_{\tau_i}$ is applied to $\pi_{\tau_{i-1}}\cdots\pi_{\tau_0}(\blds{v})$ at the $i$th step for any $i \in I_{n-1}$.
The Knuth shuffle runs in time linear in $n$ and forms an induced random permutation that is uniform \cite{Skiena:2020aa}.

Random permutations have also been studied in the context of concentration of measure in probability theory \cite{Talagrand:1995aa}. In proving Theorem \ref{thm:main}, we will need the following simple observation, of which we give a short proof for completeness.
\begin{proposition}\label{prop:lipschitz}
Let $n$ be a positive integer and take $n-1$ transpositions $\sigma_i = (n-i-1, s_i)$, $i \in I_{n-1}$, of $I_n$, where $s_i \in I_{n-i}$ for any $i \in I_{n-1}$, to form an induced permutation $\pi = \pi_{\sigma_{n-2}}\cdots\pi_{\sigma_0}$. Fix $j \in I_{n-1}$ and take another $n-1$ transpositions $\sigma_i'$, $i \in I_{n-1}$, of $I_n$, where $\sigma_i' = \sigma_i$ for $i \in I_{n-1}\setminus\{j\}$ and $\sigma_j' = (n-j-1,s_j')$ with $s_j' \in I_{n-j}\setminus\{s_j\}$, to form another induced permutation $\pi' = \pi_{\sigma_{n-2}'}\cdots\pi_{\sigma_0'}$ that only differs in the $j$th induced transposition. For any sequence $\blds{v} \in A_q^n$ of length $n$ over a set $A_q$, the Hamming distance $d_H(\pi(\blds{v}),\pi'(\blds{v}))$ between $\pi(\blds{v})$ and $\pi'(\blds{v})$ is at most $3$.
\end{proposition}
\begin{IEEEproof}
For any pair $\blds{v}, \blds{w} \in A_q^n$ and any induced permutation $\pi_{\sigma}$ for a sequence of length $n$, we have $d_H(\blds{v}, \blds{w}) = d_H(\pi_{\sigma}(\blds{v}),\pi_{\sigma}(\blds{w}))$ because $\pi_{\sigma}$ changes the positions of elements in a sequence independently of the sequence it is applied to. Note also that for any pair of transpositions $\sigma = (a,b), \sigma' = (a,c)$ of $I_n$, we have $d_H(\pi_{\sigma}(\blds{v}),\pi_{\sigma'}(\blds{v})) \leq 3$. Therefore, because $\pi = \alpha\pi_{\sigma_j}\beta$ and $\pi' = \alpha\pi_{\sigma_j'}\beta$ for some permutations $\alpha$ and $\beta$, we have $d_H(\pi(\blds{v}), \pi'(\blds{v})) \leq 3$.
\end{IEEEproof}

Simply put, Proposition \ref{prop:lipschitz} says that changing the outcome of one random variable $t_i$ in the Knuth shuffle only changes the output sequence in at most three positions. We exploit this property to invoke McDiarmid's inequality on lower tails.
\begin{theorem}[McDiarmid's lower tail \cite{McDiarmid:1989aa}]\label{thm:mcdiarmid}
Let $X_0$, \dots, $X_{n-1}$ be mutually independent random variables whose images are $\mathcal{X}_0$, \dots, $\mathcal{X}_{n-1}$, respectively.
If $f \colon \prod_{i=0}^{n-1}\mathcal{X}_i \rightarrow \mathbb{R}$ is such that for any $i \in I_n$, there exists $c_i \in \mathbb{R}$ such that for any $x_0 \in \mathcal{X}_0$, \dots, $x_{n-1} \in \mathcal{X}_{n-1}$ and any $x_i' \in \mathcal{X}_i$, it holds that
\begin{align*}
\lvert f(x_0,&\dots,x_{i-1},x_i,x_{i+1},\dots,x_{n-1}) \\
&- f(x_0,\dots,x_{i-1},x_i',x_{i+1},\dots,x_{n-1})\rvert \leq c_i,
\end{align*}
then for any $t > 0$,
\begin{align*}
\opr{Pr}(f(X_0, \dots, X_{n-1}) - \mathbb{E}(f(X_0, \dots, &X_{n-1})) \leq -t)\\
&\,\leq e^{-\frac{2t^2}{\sum_{i=0}^{n-1}c_i^2}}.
\end{align*}
\end{theorem}

McDiarmid's inequality is applicable whenever the effect of changing the outcome of one of the mutually independent random variables $X_i$ the function $f$ depends on is bounded.

\section{The Proof of Theorem \ref{thm:main}}\label{sec:proof}

In this section, we prove our main theorem by giving a linear-time randomized algorithm for constructing DSSes. For convenience, we restate Theorem \ref{thm:main} here.

\setcounter{section}{1}
\setcounter{theorem}{2}
\begin{theorem}
For any integer $q \geq 2$ and any constant $p \in (0,1)$, there exists a constant $n_p$ such that for any integer $n \geq n_p$, there exists a $q$-ary DSS$(n,\lfloor np\rfloor,\rho)$, where
\begin{align*}
\rho = n\left(1-\frac{1}{q}\right)p^2 - o(n).
\end{align*}
In particular, for any constant integer $q \geq 2$ and any constant $\delta \in \left(0,1-\frac{1}{q}\right)$,
\begin{align*}
 \lim_{n \rightarrow \infty}\frac{r_q(n,\lceil n\delta\rceil)}{n} = \sqrt{\frac{q\delta}{q-1}}.
\end{align*}
\end{theorem}
\setcounter{section}{3}
\setcounter{theorem}{0}
\begin{IEEEproof}
For fixed $q$ and $p$, write $\lfloor np \rfloor = aq+b$ by integers $q$, $a$, and $b$ with $0\leq b \leq q-1$.
Let $A_q = \{a_0,\dots,a_{q-1}\}$ be a set of cardinality $q$ and define $*$ to be a distinct element $*  \not\in A_q$.
Take an arbitrary sequence $\blds{v} = (v_0, \dots, v_{n-1})$ of length $n$ over $A_q\cup\{*\}$, where for any $a_j \in A_q$, it holds that
\begin{align*}
\lvert\{i \in I_n \mid v_{i} = a_j \}\rvert =
\begin{cases}
a+1 & \text{if $j \in I_b$},\\
a & \text{otherwise.}
\end{cases}
\end{align*}
We show that applying a uniform induced random permutation to $\blds{v}$ gives a template sequence corresponding to a desired DSS with high probability.

Take $n-1$ mutually independent random variables $t_0$, \dots, $t_{n-2}$ such that  for any $i \in I_{n-1}$ and any $j \in I_{n-i}$, it holds that $\opr{Pr}(t_i = j) = \frac{1}{n-i}$.
Consider $n-1$ random transpositions $\tau_0 = (n-1, t_0)$, $\tau_1 = (n-2, t_1)$, \dots, $\tau_{n-2} = (1, t_{n-2})$ of $I_n$ and apply $\pi_{\tau_0}$, \dots, $\pi_{\tau_{n-2}}$ to $\blds{v}$ in this order.
The randomly permuted sequence $\pi_{\tau_{n-2}}\cdots\pi_{\tau_0}(\blds{v}) = (w_0,\dots,w_{n-1})$ can be seen as the output of the Knuth shuffle applied to $\blds{v}$.

For $1 \leq t \leq n-1$, define $Y_t = \sum_{i=0}^{n-1}h(w_i,w_{i+t})$, where the function $h \colon (A_q \cup \{*\}) \times (A_q \cup \{*\}) \rightarrow \{0,1\} \subseteq \mathbb{N}_0$ is defined as in Section \ref{subsec:dss} by
\begin{align*}
h(x,y) = 
\begin{cases}
1 & \text{if $x \not= y$ and $x,y \in A_q$,}\\
0 & \text{otherwise.}
\end{cases}
\end{align*}
The template sequence $\pi_{\tau_{n-2}}\cdots\pi_{\tau_0}(\blds{v})$ corresponds to a DSS of index $\rho$ if for any integer $t$, $1 \leq t \leq n-1$, it holds that $Y_t \geq \rho$.
Note that we have $a = \frac{p}{q}n - O(1)$, $b = O(1)$, and $\lfloor np\rfloor = np - O(1)$ as $n \rightarrow \infty$.
Because the induced random permutation $\pi_{\tau_{n-2}}\cdots\pi_{\tau_0}$ can be realized by the Knuth shuffle and hence is uniform, we have
\begin{align*}
\opr{Pr}(h(w_i,w_{i+t}) = 1) &= \frac{(n-2)!}{n!}(b(a+1)(\lfloor np\rfloor-a-1)\\
&\quad\quad\quad\quad\quad \ \, +(q-b)a(\lfloor np\rfloor-a))\\
&\geq \left(1-\frac{1}{q}\right)p^2 - \Theta(n^{-1}).
\end{align*}
Hence, by linearity of expectation, we have
\begin{align*}
\mathbb{E}(Y_t) &= \sum_{i=0}^{n-1}\mathbb{E}(h(w_i,w_{i+t}))\\
&= \sum_{i=0}^{n-1}\opr{Pr}(h(w_i,w_{i+t})= 1)\\
&\geq n\left(1-\frac{1}{q}\right)p^2 - \Theta(1),
\end{align*}
that is, there exists a constant $c$ such that for all sufficiently large $n$, we have $\mathbb{E}(Y_t) \geq n\left(1-\frac{1}{q}\right)p^2 - c$.
By Proposition \ref{prop:lipschitz}, changing the outcome of one of the random variables $t_0$, \dots, $t_{n-2}$ changes the resulting sequence $\pi_{\tau_{n-2}}\cdots\pi_{\tau_0}(\blds{v})$ in at most three positions,
which implies that any change to a single outcome changes $Y_t$ by at most $6$.
Hence, by Theorem \ref{thm:mcdiarmid}, for all sufficiently large $n$, we have
\begin{align*}
\opr{Pr}\left(Y_t \leq n\left(1-\frac{1}{q}\right)p^2 - c - n^{\frac{2}{3}}\right) &\leq \opr{Pr}\left(Y_t \leq \mathbb{E}(Y_t) - n^{\frac{2}{3}}\right)\\
&\leq e^{-\frac{1}{18}n^{\frac{1}{3}}}.
\end{align*}
Let $E_t$ be the event that $Y_t > n\left(1-\frac{1}{q}\right)p^2 - c - n^{\frac{2}{3}}$. By the union bound, we have
\begin{align*}
\lim_{n \rightarrow \infty}\opr{Pr}\left(\bigcap_{t=1}^{n-1} E_t\right) &= \lim_{n \rightarrow \infty}\left(1 - \opr{Pr}\left(\bigcup_{t=1}^{n-1}\overline{E_t}\right)\right)\\
&\geq \lim_{n \rightarrow \infty}\left(1 - (n-1)e^{-\frac{1}{18}n^{\frac{1}{3}}}\right)\\
&= 1.
\end{align*}
Thus, with high probability, the induced random permutation produces a $q$-ary DSS$(n,\lfloor np\rfloor,\rho)$ with $\rho = n\left(1-\frac{1}{q}\right)p^2 - o(n)$.

It remains to show that the random DSS asymptotically achieves Levenshtein's lower bound on redundancy in Theorem \ref{thm:levenshtein}.
Fix positive $\delta$ less than $1-\frac{1}{q}$. By setting
\begin{align*}
p = \sqrt{\frac{q\delta}{q-1} + f(n)}
\end{align*}
with appropriate $f(n) = o(1)$, the above existence argument provides a $q$-ary DSS$(n,r,\lceil n\delta\rceil)$ with $r = \lfloor np\rfloor$ for all sufficiently large $n$.
Hence, we have
\begin{align*}
\lim_{n \rightarrow \infty}\frac{r_q(n,\lceil n\delta\rceil)}{n} &\leq \lim_{n \rightarrow \infty}\frac{r}{n}\\
&\leq \lim_{n \rightarrow \infty}\sqrt{\frac{q\delta}{q-1} + f(n)}\\
&=\sqrt{\frac{q\delta}{q-1}},
\end{align*}
as desired. The proof is complete.
\end{IEEEproof}

As stated in the proof above, what we proved is that applying the Knuth shuffle to an arbitrary sequence in which each symbol for synchronization appears as uniformly as possible gives a desired DSS with high probability. Because following the proof straightforwardly results in a construction algorithm that outputs a DSS in the form of a template sequence, we describe in Algorithm \ref{alg:random} an equivalent algorithm that outputs a DSS in the form of a family of subsets of $\mathbb{Z}_n$.
\begin{algorithm}[H]
    \caption{Construction for a difference system of sets}
    \label{alg:random}
    \begin{algorithmic}[1]
    \REQUIRE Integers $n$, $q \ge 2$, and real $p \in (0,1)$ with $np > 1$
    \ENSURE DSS $\{Q_0,\dots,Q_{q-1}\}$ of redundancy $\lfloor np\rfloor$ over $\mathbb{Z}_n$
    \STATE Take arbitrary disjoint subsets $Q_0, \dots, Q_{q-1}$ of $I_n$ such that $\lvert\bigcup_{i=0}^{q-1} Q_i\rvert = r$ and $\lvert Q_i \rvert = \left\lfloor \frac{r}{q}\right\rfloor$ or $\left\lfloor \frac{r}{q}\right\rfloor+1$ for any $i \in I_q$, where $r = \lfloor np\rfloor$
    \STATE Apply the Knuth shuffle to $(0,1,\dots,n-1)$ to obtain a random sequence $(v_0,\dots,v_{n-1})$
    \FOR{$i$ from $0$ to $q-1$}
    \STATE $Q_i \leftarrow \{v_j \mid j \in Q_i\}$
    \ENDFOR
    \RETURN $\{Q_0,\dots,Q_{q-1}\}$
    \end{algorithmic}
\end{algorithm}
This version also runs in time linear in the length of the corresponding self-synchronizing code. Indeed, to obtain a $q$-ary DSS of redundancy $r$ over $\mathbb{Z}_n$, we may just randomly permute $\blds{v} = (0,1,\dots,n-1)$ by the Knuth shuffle in linear time and let $Q_i$ contain consecutive $\left\lfloor \frac{r+i}{q}\right\rfloor$ elements of the permuted $\blds{v}$ by defining $Q_0$ to be the set of first consecutive $\left\lfloor \frac{r}{q} \right\rfloor$ elements, $Q_1$ the set of next consecutive $\left\lfloor \frac{r+1}{q} \right\rfloor$ elements, and so forth. McDiarmid's inequality ensures that asymptotically almost surely the index of this DSS is exactly or nearly the highest possible value.

\section{Conclusion}\label{sec:conclusion}

We determined the asymptotic behavior of the highest achievable relative index of difference systems of sets for any given constant information rate, which also determined how the highest possible information rate for a given constant relative index asymptotically behaves. We did this by giving a randomized algorithm for producing with high probability DSSes that asymptotically attain Levenshtein's lower bound on redundancy. Our algorithm runs in time linear in the length of the corresponding self-synchronizing codes and works for any desired information rate, relative comma-free index, and alphabet size, resulting in the complete resolution to both the existence and construction problems in the asymptotic sense.

Interestingly, while all approaches previously employed in the literature on DSSes had struggled to produce strong self-synchronizing codes of rate that is of practical interest \cite{Fujiwara:2013h}, our probabilistic approach works universally across all information rate and relative index. With this asymptotic solution at hand, therefore, one may also be interested in seeking more applications of DSSes beyond what is known in the literature.

The problem of detecting the phase of a data stream \cite{Wang:2017aa} provides one such previously unnoted application. Let $n \geq k \geq d \geq 1$ and $q \geq 2$ be integers.
A $q$-ary \textit{phase detection sequence} PDS$(n,k,d)$ of \textit{length} $n$, \textit{window size} $k$, and \textit{minimum distance} $d$ is a sequence $(x_0,\dots,x_{n-1}) \in A_q^n$ of length $n$ over alphabet $A_q$ of cardinality $q$ such that $\min\{d_H((x_i,\dots,x_{i+k-1}), (x_j,\dots,x_{j+k-1})) \mid i,j \in I_n, i \not= j\} = d$, where the index is understood modulo $n$, that is, the collection $\mathcal{C} = \{(x_i,\dots,x_{i+k-1}) \mid i \in I_n\}$ of $n$ contiguous subsequences of length $k$ forms an error-correcting code of minimum distance $d$ with exactly $n$ codewords. The set $\mathcal{C}$ is the \textit{codebook} of the PDS.
PDSes can be seen as \textit{robust positioning sequences} \cite{Wei:2022aa,Chee:2020aa,Berkowitz:2016aa,Bruckstein:2012aa} of special kind, so that our approach to phase detection works in the same way also for the problem of robust positioning. 

A PDS allows for transmitting timing information by inserting just one digit per packet. For instance, suppose that the receiver observes one digit at a time coming from a periodic data stream $\blds{x} = (x_i)_{i \in \mathbb{Z}}$ of period $n$, where $x_i = x_j$ for any $i \equiv j \pmod{n}$.  The receiver may start collecting digits at any position of $\blds{x}$, so that if she starts at time $t$, the collected digits form the subsequence $(x_t, \dots, x_{t+k-1})$ after $k$ units of time have elapsed. The current time $t+k-1$ modulo $n$ is called the \textit{phase}. The objective is to design a sequence of long period that allows for identifying the phase by only observing a short contiguous subsequence. A PDS$(n,k,d)$ allows for identifying the phase by observing $k$ consecutive digits under the presence of at most $\left\lfloor\frac{d-1}{2}\right\rfloor$ symbol substitution errors within any consecutive $k$ digits. For more details including the known results, see \cite{Wang:2017aa,Wei:2022aa}.

It is clear that a $q$-ary self-synchronizing code $\mathcal{D}$ of length $n$, comma-free index $\rho$, and minimum distance $d$ forms a $q$-ary PDS$(n\lvert \mathcal{D}\rvert,2n-1,d')$, where $d' \geq \min\{\rho,d\}$. Indeed, by concatenating the codewords of $\mathcal{D}$, we obtain a sequence of length $n\lvert \mathcal{D}\rvert$, where any contiguous subsequence of length $2n-1$ contains a codeword of $\mathcal{D}$, ensuring $d' \geq \min\{\rho,d\}$. We say that the $i$th codeword $\blds{d}_i \in \mathcal{D}$ placed in the PDS forms the $i$th \textit{frame} of the PDS, where $i$ is its \textit{frame number}.

While the above simple construction is suboptimal in terms of length for given window size and minimum distance, a PDS constructed from a DSS and an error-correcting code inherits favorable properties of its ingredients.
For instance, take a $q$-ary linear code $\mathcal{C}$ that admits efficient error correction and combine a DSS over $\mathbb{Z}_n$ to form a PDS. While the window size of the resulting PDS is technically $2n-1$, typically the receiver does not need to observe $2n-1$ digits to infer the phase. Indeed, as long as the collected subsequence contains a codeword of the corresponding self-synchronizing code, the receiver can locate a codeword of $\mathcal{C}$ and identify the current frame and hence the phase. Moreover, if each codeword $\blds{c}_i \in \mathcal{C}$ encodes the frame number $i$ in base $q$ using a generator matrix in systematic form \cite{MacWilliams:1977}, the current frame number $i$ can be directly read off from the information digits of the located codeword of $\mathcal{C}$ after applying efficient error correction.

As illustrated by the above simple example, difference systems of sets possess interesting combinatorial properties that lead to nontrivial applications in coding theory. It is hoped that the asymptotic resolution to their existence and construction problems given in this paper stimulates further fruitful research in this field.

\section*{Acknowledgment}
The authors thank the anonymous referees for their valuable comments. This work was supported by JSPS KAKENHI Grant Numbers JP21J00593 and JP22KJ0344 (Y.T.) and KAKENHI Grant Number JP20K11668 (Y.F.).


\end{document}